\documentclass[10pt]{article}
\usepackage{amsfonts,amsmath,amssymb,latexsym}
\usepackage{multirow}
\usepackage[longnamesfirst,round]{natbib}
\usepackage{hyperref,graphicx}
\usepackage{setspace}
\usepackage{enumitem}
\usepackage{booktabs}
\usepackage{color}

\newcommand{\nrm}{\mathcal{N}}
\newcommand{\lognrm}{\mathcal{LN}}
\newcommand{\unif}{\mathcal{U}}
\newcommand{\gam}{\mathcal{G}}
\newcommand{\lap}{\mathcal{L}}
\newcommand{\bet}{\mathcal{B}}
\newcommand{\kumar}{\mathcal{K}}
\newcommand{\tdist}{\mathcal{T}}
\newcommand{\e}{\mathbb{E}}
\newcommand{\var}{\mathbb{V}}
\newcommand{\p}{\mathbb{P}}
\newcommand{\bs}{\boldsymbol}
\newcommand{\mbf}{\mathbf}

\newcommand{\diag}{\mathop{\mathrm{diag}}}

\newcommand{\bx}{\bs{x}}

\newcommand{\by}{\bs{y}}
\newcommand{\bY}{\bs{Y}}
\newcommand{\bz}{\bs{z}}
\newcommand{\bZ}{\bs{Z}}
\newcommand{\bU}{\bs{U}}

\newcommand{\bp}{\bs{p}}

\newcommand{\bzero}{\bs{0}}

\newcommand{\bbeta}{\bs{\beta}}

\newcommand{\bomega}{\bs{\omega}}

\newcommand{\bpsi}{\bs{\psi}}

\newcommand{\btheta}{\bs{\theta}}

\newcommand{\mr}{\mbf{R}}

\newcommand{\momega}{\mbf{\Omega}}

\newcommand{\id}{\mbf{I}}

\newcommand{\meat}{\bs{\mathcal{J}}}
\newcommand{\info}{\bs{\mathcal{I}}}

\newcommand{\cml}{\text{\textsc{cml}}}
\newcommand{\dt}{\text{\textsc{dt}}}

\newcommand{\bull}{\text{\raisebox{1pt}{\scalebox{.6}{$\bullet$}}}}

\newcommand{\fct}[1]{\texttt{#1()}}
\newcommand{\proglang}[1]{\textsf{#1}}
\newcommand{\pkg}[1]{\textbf{#1}}

\author{John Hughes\,\thanks{jphughesjr@gmail.com; www.johnhughes.org}}

\title{\pkg{sklarsomega}: An \proglang{R} Package for Measuring Agreement Using Sklar's Omega Coefficient}

\begin{document}

\maketitle

\begin{abstract}
  \proglang{R} package \pkg{sklarsomega} provides tools for measuring agreement using Sklar's $\omega$ coefficient, which subsumes Krippendorff's $\alpha$ coefficient, which in turn subsumes a number of other well-known agreement coefficients. The package permits users to apply the $\omega$ methodology to nominal, ordinal, interval, or ratio scores; can accommodate any number of units, any number of coders, and missingness; and can measure intra-coder agreement, inter-coder agreement, and agreement relative to a gold standard. Classical inference is available for all levels of measurement while Bayesian inference is available for interval data and ratio data only.
\medskip

\end{abstract}

\section[Introduction: Measuring agreement in R]{Introduction: Measuring agreement in \proglang{R}}
\label{intro}

Sklar's $\omega$ \citep{omega} is a model-based alternative to Krippendorff's $\alpha$ \citep{hayes2007answering}, a well-known nonparametric measure of agreement. Although Krippendorff's $\alpha$ is intuitive, flexible, and subsumes a number of other coefficients of agreement, Sklar's $\omega$ improves upon $\alpha$ in (at least) the following ways. Sklar's $\omega$
\begin{itemize}
\item permits practitioners to simultaneously assess intra-coder agreement, inter-coder agreement, agreement with a gold standard, and, in the context of multiple scoring methods, inter-method agreement;
\item identifies the above mentioned types of agreement with intuitive, well-defined population parameters;
\item can accommodate any number of coders, any number of methods, any number of replications (per coder and/or per method), and missing values;
\item allows practitioners to use regression analysis to reveal important predictors of agreement (e.g., coder experience level, or time effects such as learning and fatigue);
\item provides complete inference, i.e., point estimation, interval estimation, diagnostics, model selection;
\item performs more robustly in the presence of unusual coders, units, or scores; and
\item permits Bayesian inference for interval or ratio scores.
\end{itemize}

In \proglang{R} \citep{Ihak:Gent:r::1996}, Krippendorff's $\alpha$ can be applied using function \fct{kripp.alpha} of package \pkg{irr} \citep{irr} and function \fct{kripp.boot} of package \pkg{kripp.boot} \citep{kripp.boot}.

\section{The agreement problem}
\label{problem}

The literature on agreement contains two broad classes of methods. Methods in the first class seek to measure agreement while also explaining disagreement---by, for example, assuming differences among coders (as in \citet{aravind2017statistical}). Although our approach permits one to use regression to explain systematic variation away from a gold standard, we are not, in general, interested in explaining disagreement. Our methodology is for measuring agreement, and so we do not typically accommodate (i.e., model) disagreement. For example, we assume that coders are exchangeable (unless multiple scoring methods are being considered, in which case we assume coder exchangeability within each method). This modeling orientation allows disagreement to count fully (up to randomness) against agreement, as desired. 

\section{Models and software}
\label{models}

The statistical model underpinning Sklar's $\omega$ is a Gaussian copula model \citep{xue2000multivariate}. We begin by specifying the most general form of the model. Then we consider special cases of the model that speak to the tasks, assumptions, and levels of measurement presented in Section~\ref{intro}.

The stochastic form of the Gaussian copula model is given by
\begin{align}
\label{gausscop}
\nonumber\bZ = (Z_1,\dots,Z_n)^\top  & \; \sim\;  \nrm(\bzero,\momega)\\
\nonumber U_i = \Phi(Z_i) & \;\sim\; \unif(0,1)\;\;\;\;\;\;\;(i=1,\dots,n)\\
Y_i = F_i^{-1}(U_i) & \;\sim\; F_i,
\end{align}
where $\momega$ is a correlation matrix, $\Phi$ is the standard Gaussian cdf, and $F_i$ is the cdf for the $i$th outcome $Y_i$. Note that $\bU=(U_1,\dots, U_n)^\top$ is a realization of the Gaussian copula, which is to say that the $U_i$ are marginally standard uniform and exhibit the Gaussian correlation structure defined by $\momega$. Since $U_i$ is standard uniform, applying the inverse probability integral transform to $U_i$ produces outcome $Y_i$ having the desired marginal distribution $F_i$.

In the form of Sklar's $\omega$ that most closely resembles Krippendorff's $\alpha$, we assume that all of the outcomes share the same marginal distribution $F$. The choice of $F$ is then determined by the level of measurement.

While Krippendorff's $\alpha$ typically employs two different metrics for nominal and ordinal outcomes, we assume the categorical distribution
\begin{align}
\label{cat}
\nonumber p_k &= \p(Y=k)\;\;\;\;\;\;(k=1,\dots,K)\\
\sum_k p_k &= 1
\end{align}
for both levels of measurement, where $K$ is the number of categories. For $K=2$, (\ref{cat}) is of course the Bernoulli distribution.

Note that when the marginal distributions are discrete (in our case, categorical), the joint distribution corresponding to (\ref{gausscop}) is uniquely defined only on the support of the marginals, and the dependence between a pair of random variables depends on the marginal distributions as well as on the copula. \citet{genest2007primer} described the implications of this and warned that, for discrete data, ``modeling and interpreting dependence through copulas is subject to caution.'' But \citeauthor{genest2007primer} go on to say that copula parameters may still be interpreted as dependence parameters, and estimation of copula parameters is often possible using fully parametric methods. It is precisely such methods that we recommend, and support in package \pkg{sklarsomega}.

For interval outcomes $F$ can be practically any continuous distribution. Our package supports the Gaussian, Laplace, Student's $t$, and gamma distributions, which we denote as $\nrm(\mu,\sigma)$, $\lap(\mu,\sigma)$, $\tdist(\nu,\mu)$, and $\gam(\alpha,\beta)$, respectively. The Laplace and $t$ distributions are useful for accommodating heavier-than-Gaussian tails, and the $t$ and gamma distributions can accommodate asymmetry.

Another possibility for continuous outcomes is to first estimate $F$ nonparametrically, and then estimate the copula parameters in a second stage. In Section~\ref{semiparametric} we will provide details regarding this approach.

Finally, two natural choices for ratio outcomes are the beta and Kumaraswamy distributions, the two-parameter versions of which are supported by our package. We denote these distributions as $\bet(\alpha,\beta)$ and $\kumar(a,b)$.

Now we turn to the copula correlation matrix $\momega$, the form of which is determined by the question(s) we seek to answer. If we wish to measure only inter-coder agreement, as is the case for Krippendorff's $\alpha$, our copula correlation matrix has a very simple structure: block diagonal, where the $i$th block corresponds to the $i$th unit $(i=1,\dots,n_u)$ and has a compound symmetry  structure. That is,
\[
\momega = \diag(\momega_i),
\]
where
\[
\momega_i = \bordermatrix{ & c_1 & c_2 & \dots & c_{n_c} \cr
c_1 & 1 & \omega  &\dots & \omega\cr
c_2 & \omega &  1 &  \dots & \omega\cr
\vdots & \vdots & \vdots  & \ddots  & \vdots\cr
c_{n_c} & \omega &  \omega & \dots  & 1
}
\]

On the scale of the outcomes, $\omega$'s interpretation depends on the marginal distribution. If the outcomes are Gaussian, $\omega$ is the Pearson correlation between $Y_{ij}$ and $Y_{ij'}$, and so the outcomes carry exactly the correlation structure codified in $\momega$. If the outcomes are non-Gaussian, the interpretation of $\omega$ (still on the scale of the outcomes) is more complicated. For example, if the outcomes are Bernoulli, $\omega$ is often called the tetrachoric correlation between those outcomes. Tetrachoric correlation is constrained by the marginal distributions. Specifically, the maximum correlation for two binary random variables is
\[
\min\left\{\sqrt{\frac{p_1(1-p_2)}{p_2(1-p_1)}},\sqrt{\frac{p_2(1-p_1)}{p_1(1-p_2)}}\right\},
\]
where $p_1$ and $p_2$ are the expectations \citep{prentice1988correlated}. More generally, the marginal distributions impose bounds, called the Fr\'{e}chet--Hoeffding bounds, on the achievable correlation \citep{Nelsen2006An-Introduction}. For most scenarios, the Fr\'{e}chet--Hoeffding bounds do not pose a problem for Sklar's $\omega$ because we typically assume that our outcomes are identically distributed, in which case the bounds are $-1$ and 1. (We do, however, impose our own lower bound of 0 on $\omega$ since we aim to measure agreement.)

In any case, $\omega$ has a uniform and intuitive interpretation for suitably transformed outcomes, irrespective of the marginal distribution. Specifically,
\[
\omega=\rho\left[\Phi^{-1}\{F(Y_{ij})\},\,\Phi^{-1}\{F(Y_{ij'})\}\right],
\]
where $\rho$ denotes Pearson's correlation and the second subscripts index the scores within the $i$th unit ($j,j'\in\{1,\dots,n_c\}:j\neq j'$). 

By changing the structure of the blocks $\momega_i$ we can use Sklar's $\omega$ to measure not only inter-coder agreement but also a number of other types of agreement. For example, should we wish to measure agreement with a gold standard, we might employ
\[
\momega_i = \bordermatrix{ & g & c_1 & c_2 & \dots & c_{n_c} \cr
g & 1 & \omega_g  & \omega_g & \dots & \omega_g\cr
c_1 & \omega_g &  1 &  \omega_c & \dots & \omega_c\cr
c_2 & \omega_g & \omega_c & 1 & \dots & \omega_c\cr
\vdots & \vdots & \vdots  & \vdots & \ddots  & \vdots\cr
c_{n_c} & \omega_g &  \omega_c & \omega_c & \dots  & 1
}.
\]
In this scheme $\omega_g$ captures agreement with the gold standard, and $\omega_c$ captures inter-coder agreement.

In a more elaborate form of this scenario, we could include a regression component in an attempt to identify important predictors of agreement with the gold standard. This could be accomplished by using a cdf to link coder-specific covariates with $\omega_g$. Then the blocks in $\momega$ might look like
\[
\momega_i = \bordermatrix{ & g & c_1 & c_2 & \dots & c_{n_c} \cr
g & 1 & \omega_{g1}  & \omega_{g2} & \dots & \omega_{gn_c}\cr
c_1 & \omega_{g1} &  1 &  \omega_c & \dots & \omega_c\cr
c_2 & \omega_{g2} & \omega_c & 1 & \dots & \omega_c\cr
\vdots & \vdots & \vdots  & \vdots & \ddots  & \vdots\cr
c_{n_c} & \omega_{gn_c} &  \omega_c & \omega_c & \dots  & 1
},
\]
where $\omega_{gj}=H(\bx_j^\top\bbeta)$, $H$ being a cdf, $\bx_j$ being a vector of covariates for coder $j$, and $\bbeta$ being regression coefficients.

For our final example we consider a complex study involving a gold standard, multiple scoring methods, multiple coders, and multiple scores per coder. In the interest of concision, suppose we have two methods, two coders per method, two scores per coder for each method, and gold standard measurements for the first method. Then $\momega_i$ is given by
\[
\momega_i = \bordermatrix{ & g_1 & c_{111} & c_{112} & c_{121} & c_{122} & c_{211} & c_{212} & c_{221} & c_{222}\cr
g_1 & 1 & \omega_{g1}  & \omega_{g1} & \omega_{g1} & \omega_{g1} & 0 & 0 & 0 & 0\cr
c_{111} & \omega_{g1} &  1 &  \omega_{11\bull} & \omega_{1\bull\bull} & \omega_{1\bull\bull} & \omega_{\bull\bull\bull} & \omega_{\bull\bull\bull} & \omega_{\bull\bull\bull} & \omega_{\bull\bull\bull}\cr
c_{112} & \omega_{g1} & \omega_{11\bull} & 1 & \omega_{1\bull\bull} & \omega_{1\bull\bull} & \omega_{\bull\bull\bull} & \omega_{\bull\bull\bull} & \omega_{\bull\bull\bull} & \omega_{\bull\bull\bull} &\cr
c_{121} & \omega_{g1} & \omega_{1\bull\bull} & \omega_{1\bull\bull} & 1 & \omega_{12\bull} & \omega_{\bull\bull\bull} & \omega_{\bull\bull\bull} & \omega_{\bull\bull\bull} & \omega_{\bull\bull\bull} &\cr
c_{122} & \omega_{g1} & \omega_{1\bull\bull} & \omega_{1\bull\bull} & \omega_{12\bull} & 1 & \omega_{\bull\bull\bull} & \omega_{\bull\bull\bull} & \omega_{\bull\bull\bull} & \omega_{\bull\bull\bull} &\cr
c_{211} & 0 & \omega_{\bull\bull\bull} & \omega_{\bull\bull\bull} & \omega_{\bull\bull\bull} & \omega_{\bull\bull\bull} &  1 & \omega_{21\bull} & \omega_{2\bull\bull} & \omega_{2\bull\bull}\cr
c_{212} & 0 & \omega_{\bull\bull\bull} & \omega_{\bull\bull\bull} & \omega_{\bull\bull\bull} & \omega_{\bull\bull\bull} & \omega_{21\bull} & 1 & \omega_{2\bull\bull} & \omega_{2\bull\bull}\cr
c_{221} & 0 & \omega_{\bull\bull\bull} & \omega_{\bull\bull\bull} & \omega_{\bull\bull\bull} & \omega_{\bull\bull\bull} & \omega_{2\bull\bull} & \omega_{2\bull\bull} & 1 & \omega_{22\bull}\cr
c_{222} & 0 & \omega_{\bull\bull\bull} & \omega_{\bull\bull\bull} & \omega_{\bull\bull\bull} & \omega_{\bull\bull\bull} & \omega_{2\bull\bull} & \omega_{2\bull\bull} & \omega_{22\bull} & 1\cr
},
\]
where the subscript $mcs$ denotes score $s$ for coder $c$ of method $m$. Thus $\omega_{g1}$ represents agreement with the gold standard for the first method, $\omega_{11\bull}$ represents intra-coder agreement for the first coder of the first method, $\omega_{12\bull}$ represents intra-coder agreement for the second coder of the first method, $\omega_{1\bull\bull}$ represents inter-coder agreement for the first method, and so on, with $\omega_{\bull\bull\bull}$ representing inter-method agreement.

Note that, for a study involving multiple methods, it may be reasonable to assume a different marginal distribution for each method. In this case, the Fr\'{e}chet--Hoeffding bounds may be relevant, and, if some marginal distributions are continuous and some are discrete, maximum likelihood inference is infeasible (see the next section for details).

\subsection{Classical inference for all levels of measurement}

When the response is continuous, i.e., when the level of measurement is interval or ratio, we recommend maximum likelihood (ML) or Bayesian inference for Sklar's $\omega$. When the marginal distribution is a categorical distribution (for nominal or ordinal level of measurement), likelihood-based inference is infeasible because the log-likelihood, having $\Theta(2^n)$ terms, is intractable for most datasets. In this case we recommend the distributional transform (DT) approximation or composite marginal likelihood (CML), depending on the number of categories. If the response is binary, composite marginal likelihood is indicated even for large samples since the DT approach tends to perform poorly for binary data. If there are three or four categories, the DT approach may perform at least adequately for larger samples, but we still favor CML for such data. For five or more categories, the DT approach performs well and yields a more accurate estimator than does the CML approach. The DT approach is also more computationally efficient than the CML approach.

\subsubsection[The method of maximum likelihood for Sklar's Omega]{The method of maximum likelihood for Sklar's $\omega$}

For correlation matrix $\momega(\bomega)$, marginal cdf $F(y\mid\bpsi)$, and marginal pdf $f(y\mid\bpsi)$, the log-likelihood of the parameters $\btheta=(\bomega^\top,\bpsi^\top)^\top$ given observations $\by$ is
\begin{align}
\label{loglik}
\ell_\textsc{ml}(\btheta\mid\by)=-\frac{1}{2}\log\vert\momega\vert-\frac{1}{2}\bz^\top(\momega^{-1}-\id)\bz+\sum_i\log f(y_i),
\end{align}
where $z_i=\Phi^{-1}\{F(y_i)\}$ and $\id$ denotes the $n\times n$ identity matrix. We obtain $\hat{\btheta}_\textsc{ml}$ by minimizing $-\ell_\textsc{ml}$. For all three approaches to inference---ML, DT, CML---we use the optimization algorithm proposed by \citet{byrd1995limited} so that $\bomega$, and perhaps some elements of $\bpsi$, can be appropriately constrained. To estimate an asymptotic confidence ellipsoid we of course use the observed Fisher information matrix, i.e., the estimate of the Hessian matrix at $\hat{\btheta}_\textsc{ml}$:
\[
\{\btheta:(\hat{\btheta}_\textsc{ml}-\btheta)^\top\,\hat{\info}_\textsc{ml}\,(\hat{\btheta}_\textsc{ml}-\btheta)\leq\chi^2_{1-\alpha,q}\},
\]
where $\hat{\info}_\textsc{ml}$ denotes the observed information and $q=\dim(\btheta)$.

Optimization of $\ell_\textsc{ml}$ is insensitive to the starting value for $\bomega$, but it can be important to choose an initial value $\bpsi_0$ for $\bpsi$ carefully. For example, if the assumed marginal family is $t$, we recommend $\bpsi_0=(\mu_0,\nu_0)^\top=(\text{med}_n,\text{mad}_n)^\top$ \citep{Serfling:2009p1313}, where $\mu$ is the noncentrality parameter, $\nu$ is the degrees of freedom, $\text{med}_n$ is the sample median, and $\text{mad}_n$ is the sample median absolute deviation from the median. For the Gaussian and Laplace distributions we use the sample mean and standard deviation. For the gamma distribution we recommend $\bpsi_0=(\alpha_0,\beta_0)^\top$, where
\begin{align*}
\alpha_0 &= \bar{Y}^2/S^2\\
\beta_0 &= \bar{Y}/S^2,
\end{align*}
for sample mean $\bar{Y}$ and sample variance $S^2$. Similarly, we provide initial values
\begin{align*}
\alpha_0 &= \bar{Y}\left\{\frac{\bar{Y}(1-\bar{Y})}{S^2}-1\right\}\\
\beta_0 &= (1-\bar{Y})\left\{\frac{\bar{Y}(1-\bar{Y})}{S^2}-1\right\}
\end{align*}
when the marginal distribution is beta. For the Kumaraswamy distribution we use $a_0=1$ and $b_0=1$. Finally, for a categorical distribution we use the empirical probabilities.

\subsubsection{The distributional transform method}

When the marginal distribution is discrete (in our case, categorical), the log-likelihood does not have the simple form given above because $z_i=\Phi^{-1}\{F(y_i)\}$ is not standard Gaussian (since $F(y_i)$ is not standard uniform if $F$ has jumps). In this case the true log-likelihood is intractable unless the sample is rather small. An appealing alternative to the true log-likelihood is an approximation based on the distributional transform.

It is well known that if $Y\sim F$ is continuous, $F(Y)$ has a standard uniform distribution. But if $Y$ is discrete, $F(Y)$ tends to be stochastically larger, and $F(Y^-)=\lim_{x\nearrow Y}F(x)$ tends to be stochastically smaller, than a standard uniform random variable. This can be remedied by stochastically ``smoothing'' $F$'s discontinuities. This technique goes at least as far back as \citet{Ferguson:1969p1279}, who used it in connection with hypothesis tests. More recently, the distributional transform has been applied in a number of other settings---see, e.g., \citet{ruschendorf1981stochastically}, \citet{burgert2006optimal}, and \citet{Ruschendorf:2009p1281}.

Let $W\sim\mathcal{U}(0,1)$, and suppose that $Y\sim F$ and is independent of $W$. Then the distributional transform
\[
G(W,Y)=WF(Y^-)+(1-W)F(Y)
\]
follows a standard uniform distribution, and $F^{-1}\{G(W,Y)\}$ follows the same distribution as $Y$.

\citet{Kazianka:2010p941} suggested approximating $G(W,Y)$ by replacing it with its expectation with respect to $W$:
\begin{align*}
G(W,Y) &\approx \e_W G(W,Y)\\
&= \e_W\{WF(Y^-)+(1-W)F(Y)\}\\
&= \e_W WF(Y^-) + \e_W (1-W)F(Y)\\
&= F(Y^-)\e_W W + F(Y)\e_W(1-W)\\
&= \frac{F(Y^-) + F(Y)}{2}.
\end{align*}
To construct the approximate log-likelihood for Sklar's $\omega$, we replace $F(y_i)$ in (\ref{loglik}) with
\[
\frac{F(y_i^-) + F(y_i)}{2}.
\]
If the distribution has integer support, this becomes
\[
\frac{F(y_i-1) + F(y_i)}{2}.
\]

This approximation is crude, but it performs well as long as the discrete distribution in question has a sufficiently large variance \citep{Kazianka2013}. For Sklar's $\omega$, we recommend using the DT approach when the scores fall into five or more categories.

Since the DT-based objective function is misspecified, using $\hat{\info}_\textsc{dt}$ alone leads to optimistic inference unless the number of categories is large. This can be overcome by using a sandwich estimator \citep{godambe1960optimum} or by doing a bootstrap \citep{davison1997bootstrap}.

\subsubsection{Composite marginal likelihood}

For nominal or ordinal outcomes falling into a small number of categories, we recommend a composite marginal likelihood \citep{Lindsay:1988p1155,varin2008composite} approach to inference. Our objective function comprises pairwise likelihoods (which implies the assumption that any two pairs of outcomes are independent). Specifically, we work with log composite likelihood
\begin{align*}
\ell_\cml(\btheta\mid\by) &= \mathop{\mathop{\sum_{i\in\{1,\dots,n-1\}}}_{j\in\{i+1,\dots,n\}}}_{\momega_{ij}\neq 0}\log\left\{\sum_{j_1=0}^1\sum_{j_2=0}^1(-1)^k\Phi_{\momega^{ij}}(z_{ij_1},z_{jj_2})\right\},
\end{align*}
where $k=j_1+j_2$, $\Phi_{\momega^{ij}}$ denotes the cdf for the bivariate Gaussian distribution with mean zero and correlation matrix
\[
\momega^{ij}=\begin{pmatrix}1&\momega_{ij}\\\momega_{ij}&1\end{pmatrix},
\]
$z_{\bull 0}=\Phi^{-1}\{F(y_\bull)\}$, and $z_{\bull 1}=\Phi^{-1}\{F(y_\bull-1)\}$. Since this objective function, too, is misspecified, bootstrapping or sandwich estimation is necessary.

\subsubsection{Sandwich estimation for the DT and CML procedures}
\label{sandwich}

As we mentioned above, the DT and CML objective functions are misspecified, and so the asymptotic covariance matrices of $\hat{\btheta}_\dt$ and $\hat{\btheta}_\cml$ have sandwich forms \citep{godambe1960optimum,Geyer2005Le-Cam-Made-Sim}. Specifically, we have
\begin{align*}
\sqrt{n}(\hat{\btheta}_\cml-\btheta) &\;\;\Rightarrow_n\;\; \nrm\{\bzero,\;\info_\cml^{-1}(\btheta)\meat_\cml(\btheta)\info_\cml^{-1}(\btheta)\}\\
\sqrt{n}(\hat{\btheta}_\dt-\btheta) &\;\;\Rightarrow_n\;\; \nrm\{\bzero,\;\info_\dt^{-1}(\btheta)\meat_\dt(\btheta)\info_\dt^{-1}(\btheta)\},
\end{align*}
where $\info_\bull$ is the appropriate Fisher information matrix and $\meat_\bull$ is the variance of the score:
\[
\meat_\bull(\btheta)=\var\nabla\ell_\bull(\btheta\mid\bY).
\]
We recommend that $\meat_\bull$ be estimated using a parametric bootstrap, i.e, our estimator of $\meat_\bull$ is
\[
\hat{\meat}_\bull(\btheta)=\frac{1}{n_b}\sum_{j=1}^{n_b}\nabla\nabla^\prime\ell_\bull(\hat{\btheta}_\bull\mid\bY^{(j)}),
\]
where $n_b$ is the bootstrap sample size and the $\bY^{(j)}$ are datasets simulated from our model at $\btheta=\hat{\btheta}_\bull$. This approach performs well and is considerably more efficient computationally than a ``full'' bootstrap (it is much faster to approximate the score than to optimize the objective function). What is more, $\hat{\meat}_\bull(\btheta)$ is accurate for small bootstrap sample sizes (100 in our simulations). Package \pkg{sklarsomega} makes the procedure even more efficient through parallelization.

\subsection{A two-stage semiparametric approach for continuous measurements}
\label{semiparametric}

If the sample size is large enough, a two-stage semiparametric method (SMP) may be used. In the first stage one estimates $F$ nonparametrically. The empirical distribution function $\hat{F}_n(y)=n^{-1}\sum_i 1\{Y_i\leq y\}$ is a natural choice for our estimator of $F$, but other sensible choices exist. For example, one might employ the Winsorized estimator
\begin{align*}
\tilde{F}_n(y)=\begin{cases}
\epsilon_n &\text{if }\,\hat{F}_n(y)<\epsilon_n\\
\hat{F}_n(y) &\text{if }\,\epsilon_n\leq\hat{F}_n(y)\leq 1-\epsilon_n\\
1-\epsilon_n &\text{if }\,\hat{F}_n(y)>1-\epsilon_n,
\end{cases}
\end{align*}
where $\epsilon_n$ is a truncation parameter \citep{klaassen1997efficient,liu2009nonparanormal}. A third possibility is a smoothed empirical distribution function
\[
\breve{F}_n(y)=\frac{1}{n}\sum_iK_n(y-Y_i),
\]
where $K_n$ is a kernel \citep{smoothedecdf}.

Armed with an estimate of $F$---$\hat{F}_n$, say---we compute $\hat{\bz}$, where $\hat{z}_i=\Phi^{-1}\{\hat{F}_n(y_i)\}$, and optimize
\begin{align*}
\ell_\textsc{ml}(\bomega\mid\hat{\bz})=-\frac{1}{2}\log\vert\momega\vert-\frac{1}{2}\hat{\bz}^\top\momega^{-1}\hat{\bz}
\end{align*}
to obtain $\hat{\bomega}$. This approach is advantageous when the marginal distribution is complicated, but has the drawback that uncertainty regarding the marginal distribution is not reflected in the (ML) estimate of $\hat{\bomega}$'s variance. This deficiency can be avoided by using a bootstrap sample $\{\hat{\bomega}^*_1,\dots,\hat{\bomega}^*_{n_b}\}$, the $j$th element of which can be generated by (1) simulating $\bU^*_j$ from the copula at $\bomega=\hat{\bomega}$; (2) computing a new response $\bY^*_j$ as $Y^*_{ji}=\hat{F}^{-1}_n(U^*_{ji})\;\;(i=1,\dots,n)$, where $\hat{F}^{-1}_n(p)$ is the empirical quantile function; and (3) applying the estimation procedure to $\bY^*_j$. We compute sample quantiles using the median-unbiased approach recommended by \citet{quantiles}. It is best to compute the bootstrap interval using the Gaussian method since that interval tends to have the desired coverage rate while the quantile method tends to produce an interval that is too narrow. This is because the upper-quantile estimator is inaccurate while the bootstrap estimator of $\var\hat{\bomega}$ is rather more accurate. To get adequate performance using the quantile method, a much larger sample size is required.

Although this approach may be necessary when the marginal distribution does not appear to take a familiar form, two-stage estimation does have a significant drawback, even for larger samples. If agreement is at least fair (see the introduction to Section~\ref{illustrations} for information regarding interpretation of agreement coefficients), dependence may be sufficient to pull the empirical marginal distribution away from the true marginal distribution. In such cases, simultaneous estimation of the marginal distribution and the copula should perform better. Development of such a method will be the aim of a future project.

\subsection{Bayesian inference for interval or ratio scores}
\label{bayes}

Since the Sklar's $\omega$ likelihood is not available in the case of nominal or ordinal scores, {\em true} Bayesian inference is infeasible for those levels of measurement. It is possible, however, to do {\em pseudo}-Bayesian inference for discrete scores. This entails using the appropriate CML or DT-based objective function in place of the likelihood. Although sound theory supports this approach \citep{ribatet2012bayesian}, package \pkg{sklarsomega} does not support pseudo-Bayesian inference, for two reasons. First, pseudo-Bayesian inference requires a curvature correction because both the CML and the DT-based objective functions have too large a curvature relative to the true likelihood; unfortunately, the curvature adjustment is based on a time-consuming frequentist procedure. Second, we have no reason to suspect that the (curvature-adjusted) pseudo-posterior will have (at least approximately) the same shape as the true posterior.

Via function \fct{sklars.omega.bayes}, package \pkg{sklarsomega} does support Bayesian inference for interval or ratio scores. The function's signature appears below.
\begin{verbatim}
sklars.omega.bayes(data, level = c("interval", "ratio"), verbose = FALSE,
  control = list())
\end{verbatim}

As we mentioned above, package \pkg{sklarsomega} currently supports gamma, Gaussian, Laplace, and $t$ marginal distributions for interval outcomes; and beta and Kumaraswamy marginals for ratio outcomes.

The Sklar's $\omega$ posterior is given by
\[
\pi(\btheta\mid\by) \propto L(\btheta\mid\by)p(\btheta),
\]
where $p(\cdot)$ denotes a prior distribution and
\[
L(\btheta\mid\by)=\frac{\frac{1}{\vert\momega\vert^{1/2}}\exp[-\frac{1}{2}\bz^\top\{\momega(\bomega)^{-1}-\id\}\bz]}{\prod_i f(y_i\mid\bpsi)}.
\]
In the interest of striking a sensible balance between flexibility and usability, we do not permit the user to specify $p(\btheta)$. Instead, we assign an independent, noninformative prior to each element of $\btheta$---i.e., $p(\btheta)=p(\bomega)p(\bpsi)=\{\prod_{k=1}^mp(\omega_k)\}p(\psi_1)p(\psi_2)$. The prior for $\omega_k\;\;(k=1,\dots,m)$ is standard uniform. Each of $\alpha$, $\beta$, $\sigma$, $\nu$, $a$, and $b$ is given a $\gam(0.01,0.01)$ prior distribution. And the prior for $\mu$ is Gaussian with mean zero and standard deviation 1,000.

As for sampling, we use a Gaussian random walk for each parameter, and transform when necessary. The user can control the acceptance rates by adjusting the standard deviations of the perturbations, which can be supplied to function \fct{sklars.omega.bayes} via argument \texttt{control}. Consider parameter $\alpha$, for example. To generate a proposal for $\alpha$, we begin by drawing $\eta^*=\eta+\nrm(0,\sigma_1)$, where $\eta$ was obtained during the previous iteration. Then we take $\alpha^*=\exp(\eta^*)$, which of course yields a log-normal proposal (necessitating the inclusion of the ratio $\lognrm\{\alpha;\log(\alpha^*),\sigma_1\}/\lognrm\{\alpha^*;\log(\alpha),\sigma_1\}$ in the Metropolis--Hastings acceptance probability). The proposal standard deviation $\sigma_1$ can be set using the syntax \texttt{control = list(sigma.1 = 0.2)}, for example. This proposal scheme is employed for all of the non-negative parameters, with $\sigma_1$ the tuning parameter for $\alpha$, $\nu$, and $a$; and $\sigma_2$ the tuning parameter for $\beta$, $\sigma$, and $b$. Again, these standard deviations can be set straightforwardly in the function call: \texttt{control = list(sigma.1 = 0.2, sigma.2 = 0.3)}; or they can be omitted, in which case they default to the value 0.1.

Updates for the $\mu$ chain take the form of a Gaussian random walk: $\mu^*=\mu+\nrm(0,\sigma_j)$, where $j=1$ if the marginal distribution is Gaussian or Laplace, or $j=2$ if the marginal distribution is $\tdist(\nu,\mu)$. The acceptance rate can be modulated via control parameter \texttt{sigma.1} (for a $\nrm$ or $\lap$ marginal) or \texttt{sigma.2} (for a $\tdist$ marginal).

Although we propose values for the $\omega_k$ independently, we accept or reject those proposals jointly so that $\vert\momega\vert$ and $\momega^{-1}$ need not be computed too frequently. Each proposal begins with a Gaussian random step, $\eta_k^*=\eta_k+\nrm(0,\sigma_{\omega_k})$. Then we apply the logistic function to map into the unit interval: $\omega_k^*=\exp(\eta_k^*)/\{1+\exp(\eta_k^*)\}$. This of course requires us to include the Jacobian $\exp(\eta_k^*)/\{1+\exp(\eta_k^*)\}^2$ in the Metropolis--Hastings acceptance probability. The acceptance rates can be adjusted by passing a vector of proposal standard deviations to \fct{sklars.omega.bayes}, e.g., \texttt{control = list(sigma.omega = c(0.1, 0.1, 0.3))} (in the case of $\dim(\bomega)=3$).

The remaining control parameters are \texttt{dist} (for selecting the marginal distribution), \texttt{minit}, \texttt{maxit}, and \texttt{tol}. Function \fct{sklars.omega.bayes} draws at least \texttt{minit} posterior samples. We use 1,000 as the default, and minimum, value for \texttt{minit}. Similarly, \fct{sklars.omega.bayes} draws at most \texttt{maxit} samples, with 10,000 as the default value.

Since the Markov chain tends to mix well, between 1,000 and 10,000 samples are usually sufficient for obtaining stable estimates (of posterior means and of DIC \citep{spiegelhalter2002bayesian}). We recommend using the fixed-width method \citep{Flegal:2008p1285} for determining when to stop sampling, and we provide control parameter \texttt{tol} for this purpose. In the fixed-width approach, one chooses a small positive threshold $\epsilon$ and terminates sampling when all estimated coefficients of variation are smaller than said threshold, where the estimated coefficient of variation for parameter $\theta_j$ is $\widehat{\text{cv}}_j=\text{mcse}(\hat{\theta}_j)/\vert\hat{\theta}_j\vert$, with `mcse' denoting Monte Carlo standard error. That is, sampling terminates when $\widehat{\text{cv}}_j<\epsilon$ for all $j\in\{1,\dots,\dim(\btheta)\}$. The user can set the threshold value via control parameter \texttt{tol}, which defaults to 0.1. In the interest of computational efficiency, function \fct{sklars.omega.bayes} computes Monte Carlo standard errors (using package \pkg{mcmcse}) only every \texttt{minit} samples.



\section{Illustrations}
\label{illustrations}

Here we illustrate the use of \pkg{sklarsomega} by applying Sklar's $\omega$ to a couple of datasets. Although our understanding of the agreement problem aligns with that of Krippendorff's $\alpha$ and other related measures (e.g., Spearman's $\rho$, Cohen's $\kappa$, Scott's $\pi$ \citep{spearman,cohen,scott}), we shall adopt a subtler interpretation of the results. According to \citet{krippendorff2012content}, social scientists often feel justified in relying on data for which agreement is at or above 0.8, drawing tentative conclusions from data for which agreement is at or above 2/3 but less than 0.8, and discarding data for which agreement is less than 2/3. We use the following interpretations instead (Table~\ref{tab:interpret}), and suggest---as do Krippendorff and others \citep{artstein2008inter,landiskoch}---that an appropriate reliability threshold may be context dependent.

\begin{table}[h]
\centering
\begin{tabular}{cl}
Range of Agreement & Interpretation\\\hline
$\phantom{0.2<\;}\omega\leq 0.2$ & Slight Agreement\\
$0.2<\omega\leq 0.4$ & Fair Agreement\\
$0.4<\omega\leq 0.6$ & Moderate Agreement\\
$0.6<\omega\leq 0.8$ & Substantial Agreement\\
$\phantom{0.2<\;}\omega>0.8$ & Near-Perfect Agreement\\
\end{tabular}
\caption{Guidelines for interpreting values of an agreement coefficient.}
\label{tab:interpret}
\end{table}

\subsection{Nominal data analyzed previously by Krippendorff}

Consider the following data, which appear in \citep{krippendorff2013}. These are nominal values (in $\{1,\dots,5\}$) for twelve units and four coders. The dots represent missing values.

\begin{figure}[h]
   \centering
   \begin{tabular}{ccccccccccccc}
   & $u_1$ &  $u_2$ & $u_3$ & $u_4$ & $u_5$ & $u_6$ & $u_7$ & $u_8$ & $u_9$ & $u_{10}$ & $u_{11}$ & $u_{12}$\vspace{2ex}\\
   $c_1$ & 1 & 2 & 3 & 3 & 2 & 1 & 4 & 1 & 2 & \bull & \bull & \bull\\
   $c_2$ & 1 & 2 & 3 & 3 & 2 & 2 & 4 & 1 & 2 & 5 & \bull & 3\\
   $c_3$ & \bull & 3 & 3 & 3 & 2 & 3 & 4 & 2 & 2 & 5 & 1 & \bull\\
   $c_4$ & 1 & 2 & 3 & 3 & 2 & 4 & 4 & 1 & 2 & 5 & 1 & \bull
   \end{tabular}
   \caption{Some example nominal outcomes for twelve units and four coders, with seven missing values.}
   \label{fig:nominal}
\end{figure}

Note that all columns save the sixth are constant or nearly so. This suggests near-perfect agreement, yet a Krippendorff's $\alpha$ analysis of these data leads to a weaker conclusion. Specifically, using the discrete metric $d(x,y)=1\{x\neq y\}$ yields $\hat{\alpha}=0.74$ and bootstrap 95\% confidence interval (0.39, 1.00). (We used a bootstrap sample size of $n_b=$ 1,000, which yielded Monte Carlo standard errors (MCSE) \citep{Flegal:2008p1285} smaller than 0.001.) This point estimate indicates merely substantial agreement, and the interval implies that these data are consistent with agreement ranging from moderate to nearly perfect.

Now we apply our methodology, first loading package \pkg{sklarsomega}.

\begin{verbatim}
R> library(sklarsomega)

sklarsomega: Measuring Agreement Using Sklar's Omega Coefficient
Version 2.0 created on 2018-06-18.
Copyright (c) 2018 John Hughes
For citation information, type citation("sklarsomega").
Type help(package = sklarsomega) to get started.
\end{verbatim}

Now we create the dataset as a matrix. Then we supply appropriate column names. This is a necessary step since package function \fct{build.R} uses the column names to create the copula correlation matrix (which we denoted as $\momega$ above but is denoted as $\mr$ in the package documentation).

\begin{verbatim}
R> data = matrix(c(1,2,3,3,2,1,4,1,2,NA,NA,NA,
+                  1,2,3,3,2,2,4,1,2,5,NA,3,
+                  NA,3,3,3,2,3,4,2,2,5,1,NA,
+                  1,2,3,3,2,4,4,1,2,5,1,NA), 12, 4)
R> colnames(data) = c("c.1.1", "c.2.1", "c.3.1", "c.4.1")
R> data

      c.1.1 c.2.1 c.3.1 c.4.1
 [1,]     1     1    NA     1
 [2,]     2     2     3     2
 [3,]     3     3     3     3
 [4,]     3     3     3     3
 [5,]     2     2     2     2
 [6,]     1     2     3     4
 [7,]     4     4     4     4
 [8,]     1     1     2     1
 [9,]     2     2     2     2
[10,]    NA     5     5     5
[11,]    NA    NA     1     1
[12,]    NA     3    NA    NA
\end{verbatim}

Function \fct{check.colnames} can be used to perform a (rudimentary) check of the column names. This function returns a list comprising exactly two elements, \texttt{success} and \texttt{cols}. The former is a boolean that indicates appropriateness or inappropriateness of the column names. If the column names are appropriate, element \texttt{cols} is equal to \texttt{NULL}. Otherwise, \texttt{col} contains the numbers of the columns that failed the check.

\begin{verbatim}
R> (check.colnames(data))

$success
[1] TRUE

$cols
NULL
\end{verbatim}

In the next example we introduce errors for columns 1 and 4. Then we provide column names that pass the check but are illogical. Finally, we revert to the correct names.

\begin{verbatim}
R> colnames(data) = c("c.a.1", "c.2.1", "c.3.1", "C.4.1")
R> (check.colnames(data))

$success
[1] FALSE

$cols
[1] 1 4

R> colnames(data) = c("g", "c.2.1", "c.1.47", "c.2.1")
R> (check.colnames(data))

$success
[1] TRUE

$cols
NULL

R> colnames(data) = c("c.1.1", "c.2.1", "c.3.1", "c.4.1")
\end{verbatim}

Now we create the copula correlation matrix, display the portion of the matrix that corresponds to the first two units, and verify that the matrix contains exactly one parameter, namely, the inter-coder agreement coefficient (which has the dummy value 0.1).

\begin{verbatim}
R> temp = build.R(data)
R> names(temp)

[1] "R"      "onames"

R> R = temp$R
R> R[1:7, 1:7]

     [,1] [,2] [,3] [,4] [,5] [,6] [,7]
[1,]  1.0  0.1  0.1  0.0  0.0  0.0  0.0
[2,]  0.1  1.0  0.1  0.0  0.0  0.0  0.0
[3,]  0.1  0.1  1.0  0.0  0.0  0.0  0.0
[4,]  0.0  0.0  0.0  1.0  0.1  0.1  0.1
[5,]  0.0  0.0  0.0  0.1  1.0  0.1  0.1
[6,]  0.0  0.0  0.0  0.1  0.1  1.0  0.1
[7,]  0.0  0.0  0.0  0.1  0.1  0.1  1.0

R> temp$onames

[1] "inter"
\end{verbatim}

Next we apply Sklar's $\omega$ for nominal data. Since there are five categories, function \fct{sklars.omega} automatically selects the DT approach. We do a full bootstrap with sample size $n_b=$ 1,000. We do the bootstrapping in embarrassingly parallel fashion, using six CPU cores on the local machine (hence \texttt{type} is equal to \texttt{"SOCK"}). Since we set \texttt{verbose} equal to \texttt{TRUE}, a progress bar is shown. The computation took 19 minutes and 21 seconds.

\begin{verbatim}
R> set.seed(99)
R> fit = sklars.omega(data, level = "nominal", confint = "bootstrap",
+                     verbose = TRUE, control = list(bootit = 1000,
+                     parallel = TRUE, nodes = 6))

Control parameter 'type' must be "SOCK", "PVM", "MPI", or "NWS". Setting it to "SOCK".

   |++++++++++++++++++++++++++++++++++++++++++++++++++| 100% elapsed = 19m 21s
\end{verbatim}

One can view a summary by passing the fit object to function \fct{summary.sklarsomega}.

\begin{verbatim}
R> summary(fit)

Call:

sklars.omega(data = data, level = "nominal", confint = "bootstrap", 
    verbose = TRUE, control = list(bootit = 1000, parallel = TRUE, 
        nodes = 6))

Convergence:

Optimization converged at -40.42 after 31 iterations.

Control parameters:
                    
bootit   1000       
parallel TRUE       
nodes    6          
dist     categorical
type     SOCK       
                    
Coefficients:

      Estimate    Lower  Upper
inter  0.89420  0.77530 1.0130
p1     0.25170  0.02420 0.4792
p2     0.24070  0.09078 0.3907
p3     0.22740  0.07583 0.3790
p4     0.18880  0.03522 0.3424
p5     0.09136 -0.06572 0.2484
\end{verbatim}

Our method produced $\hat{\omega}=0.89$ and $\omega\in(0.78, 1.00)$, which indicate near-perfect agreement and at least substantial agreement, respectively. And our approach, being model based, furnishes us with estimated probabilities for the marginal categorical distribution of the response:
\[
\hat{\bp}=(\hat{p}_1,\hat{p}_2,\hat{p}_3,\hat{p}_4,\hat{p}_5)^\top=(0.25, 0.24, 0.23, 0.19, 0.09)^\top.
\]
Because we estimated $\omega$ and $\bp$ simultaneously, our estimate of $\bp$ differs substantially from the empirical probabilities, which are 0.22, 0.32, 0.27, 0.12, and 0.07, respectively.

Now we compute asymptotic (sandwich) intervals. This requires a much shorter running time.

\begin{verbatim}
R> set.seed(12)
R> fit = sklars.omega(data, level = "nominal", confint = "asymptotic",
+                     verbose = TRUE, control = list(bootit = 1000,
+                     parallel = TRUE, nodes = 6))

Control parameter 'type' must be "SOCK", "PVM", "MPI", or "NWS". Setting it to "SOCK".

   |++++++++++++++++++++++++++++++++++++++++++++++++++| 100% elapsed = 02m 03s

R> summary(fit)

Call:

sklars.omega(data = data, level = "nominal", confint = "asymptotic", 
    verbose = TRUE, control = list(bootit = 1000, parallel = TRUE, 
        nodes = 6))

Convergence:

Optimization converged at -40.42 after 31 iterations.

Control parameters:
                    
bootit   1000       
parallel TRUE       
nodes    6          
dist     categorical
type     SOCK       
                    
Coefficients:

      Estimate    Lower  Upper
inter  0.89420  0.76270 1.0260
p1     0.25170  0.01703 0.4864
p2     0.24070  0.01796 0.4635
p3     0.22740  0.04792 0.4069
p4     0.18880 -0.06715 0.4447
p5     0.09136 -0.16860 0.3513
\end{verbatim}

The marked difference between our results and those obtained for Krippendorff's $\alpha$ can be attributed largely to the codes for the sixth unit. The relevant influence statistics are
\[
\delta_{\alpha}(\bull,-6)=\frac{\vert\hat{\alpha}_{\bull,-6}-\hat{\alpha}\vert}{\hat{\alpha}}=0.15
\]
and
\[
\delta_{\omega}(\bull,-6)=\frac{\vert\hat{\omega}_{\bull,-6}-\hat{\omega}\vert}{\hat{\omega}}=0.09,
\]
where the notation ``$\bull,-6$'' indicates that all rows are retained and column 6 is left out. And so we see that column 6 exerts 2/3 more influence on $\hat{\alpha}$ than it does on $\hat{\omega}$. Since $\hat{\alpha}_{\bull,-6}=0.85$, inclusion of column 6 draws us away from what seems to be the correct conclusion for these data.

Influence (DFBETA \citep{Young2017Handbook-of-Reg}) statistics can be obtained using the package's \fct{influence} function, as illustrated below. Here we investigate the influence of units 6 and 11, and of coders 2 and 3.

\begin{verbatim}
R> (inf = influence(fit, units = c(6, 11), coders = c(2, 3)))

$dfbeta.units
         inter         p1           p2          p3          p4           p5
6  -0.07914843 0.03438538  0.052599491 -0.05540904 -0.05820757  0.026631732
11  0.01096758 0.04546670 -0.007630807 -0.01626192 -0.01514173 -0.006432246

$dfbeta.coders
          inter           p1           p2          p3         p4          p5
2  0.0579843781 -0.002743713  0.002974195 -0.02730064 0.01105672  0.01601343
3 -0.0008664934 -0.006572821 -0.048168128  0.05659853 0.02149364 -0.02335122
\end{verbatim}

We conclude this illustration by simulating a dataset from the fitted model and then converting the resulting vector to matrix form for display. Note that row 12 was removed since, having only one score, it carries no information about $\omega$.

\begin{verbatim}
R> sim = simulate(fit, seed = 42)
R> data.sim = t(fit$data)
R> data.sim[! is.na(data.sim)] = sim[, 1]
R> data.sim = t(data.sim)
R> data.sim

      c.1.1 c.2.1 c.3.1 c.4.1
 [1,]     3     4    NA     3
 [2,]     3     2     2     2
 [3,]     3     3     3     3
 [4,]     1     1     1     2
 [5,]     3     4     3     4
 [6,]     5     4     4     4
 [7,]     2     2     1     2
 [8,]     4     4     4     4
 [9,]     1     2     1     1
[10,]    NA     1     2     2
[11,]    NA    NA     2     2
\end{verbatim}

\subsection{Interval data from an imaging study of hip cartilage}

\begin{figure}[h]
   \centering
   \includegraphics[scale=.25]{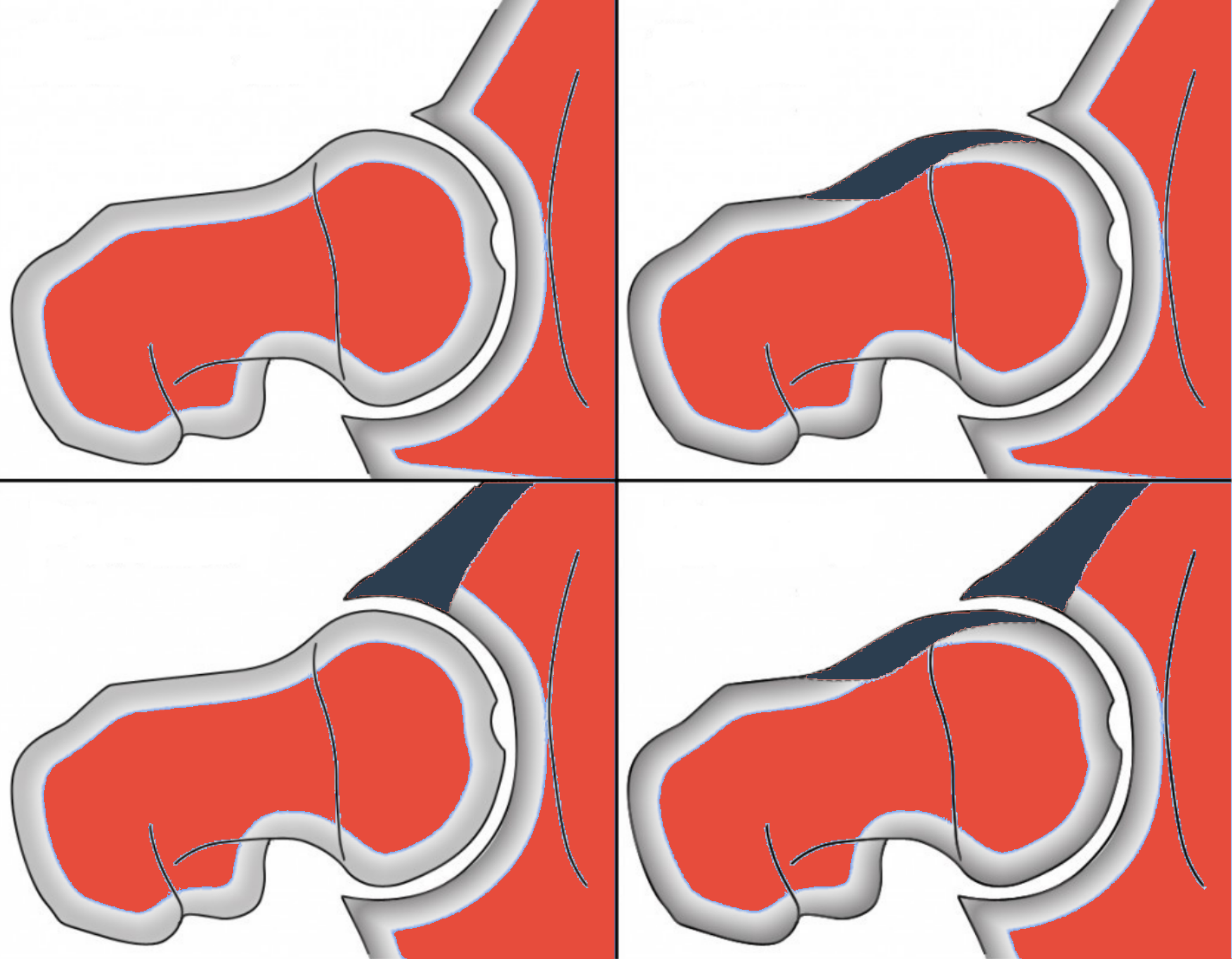}
   \caption{An illustration of femoroacetabular impingement (FAI). Top left: normal hip joint. Top right: cam type FAI. Bottom left: pincer type FAI. Bottom right: mixed type.}
   \label{fig:fai}
\end{figure}

The data for this example, some of which appear in Figure~\ref{fig:interval}, are 323 pairs of T2* relaxation times (a magnetic resonance quantity) for femoral cartilage \citep{nissi2015t2} in patients with femoroacetabular impingement (Figure~\ref{fig:fai}), a hip condition that can lead to osteoarthritis. One measurement was taken when a contrast agent was present in the tissue, and the other measurement was taken in the absence of the agent. The aim of the study was to determine whether raw and contrast-enhanced T2* measurements agree closely enough to be interchangeable for the purpose of quantitatively assessing cartilage health.

\begin{figure}[h]
   \centering
   \begin{tabular}{cccccccccccc}
   & $u_1$ &  $u_2$ & $u_3$ & $u_4$ & $u_5$ & $\dots$ & $u_{319}$ & $u_{320}$ & $u_{321}$ & $u_{322}$ & $u_{323}$\vspace{2ex}\\
   $c_1$ & 27.3 & 28.5 & 29.1 & 31.2 & 33.0 & $\dots$ & 19.7 & 21.9 & 17.7 & 22.0 & 19.5\\
   $c_2$ & 27.8 & 25.9 & 19.5 & 27.8 & 26.6 & $\dots$ & 18.3 & 23.1 & 18.0 & 25.7 & 21.7
   \end{tabular}
   \caption{Raw and contrast-enhanced T2* values for femoral cartilage.}
   \label{fig:interval}
\end{figure}

We will carry out both maximum likelihood and Bayesian analyses for (a subset of) these data, assuming $\lap(\mu,\sigma)$ and $\tdist(\nu,\mu)$ marginal distributions.

First we load the cartilage data, which are included in the package, and apply the method of maximum likelihood for a Laplace marginal distribution. The running time is just over one second.

\begin{verbatim}
R> data(cartilage)
R> data = as.matrix(cartilage)[1:100, ]
R> colnames(data) = c("c.1.1", "c.2.1")
R> fit1 = sklars.omega(data, level = "interval", confint = "asymptotic",
+                      control = list(dist = "laplace"))
R> summary(fit1)

Call:

sklars.omega(data = data, level = "interval", confint = "asymptotic", 
    control = list(dist = "laplace"))

Convergence:

Optimization converged at -593.5 after 18 iterations.

Control parameters:
            
dist laplace
            
Coefficients:

      Estimate   Lower   Upper
inter   0.8077  0.7379  0.8776
mu     26.5100 26.3500 26.6700
sigma   4.7090  3.8630  5.5550

AIC: 1193 
BIC: 1203
\end{verbatim}

We see that $\hat{\btheta}=(\hat{\omega},\hat{\mu},\hat{\sigma})^\top=(0.81, 26.51, 4.71)^\top$. This suggests that the contrast-enhanced T2* values agree nearly perfectly with their raw counterparts.

Now we repeat the analysis for a $t$ marginal distribution.

\begin{verbatim}
R> fit2 = sklars.omega(data, level = "interval", confint = "asymptotic",
+                      control = list(dist = "t"))
R> summary(fit2)

Call:

sklars.omega(data = data, level = "interval", confint = "asymptotic", 
    control = list(dist = "t"))

Convergence:

Optimization converged at -608.7 after 18 iterations.

Control parameters:
      
dist t
      
Coefficients:

      Estimate   Lower   Upper
inter   0.8701  0.8224  0.9179
nu      7.0280  5.2130  8.8430
mu     23.4400 22.2400 24.6400

AIC: 1223 
BIC: 1233 
\end{verbatim}

This led to a considerably larger value for $\hat{\omega}$ and, given the two confidence intervals, a stronger conclusion for these data. But we must select the Laplace model since that model yielded much smaller values of AIC and BIC. In fact, the model probability \citep{burnham2011aic} is near zero.

\begin{verbatim}
R> aic = c(AIC(fit1), AIC(fit2))
R> (pr = exp((min(aic) - max(aic)) / 2))

[1] 2.516706e-07
\end{verbatim}

Finally, we apply the Bayesian methodology described in Section~\ref{bayes}. We begin by assuming a Laplace marginal model once again.

\begin{verbatim}
R> set.seed(111111)
R> fit1 = sklars.omega.bayes(data, verbose = FALSE,
+                            control = list(dist = "laplace", minit = 1000,
+                            maxit = 5000, tol = 0.01, sigma.1 = 1, sigma.2 = 0.1,
+                            sigma.omega = 0.2))
R> summary(fit1)

Call:

sklars.omega.bayes(data = data, verbose = FALSE, control = list(dist = "laplace", 
    minit = 1000, maxit = 5000, tol = 0.01, sigma.1 = 1, sigma.2 = 0.1, 
    sigma.omega = 0.2))

Number of posterior samples: 4000 

Control parameters:
                   
dist        laplace
minit       1000   
maxit       5000   
tol         0.01   
sigma.1     1      
sigma.2     0.1    
sigma.omega 0.2    
                   
Coefficients:

      Estimate   Lower   Upper     MCSE
inter   0.8079  0.7366  0.8695 0.002111
mu     26.4600 25.7100 27.1400 0.011310
sigma   4.7990  3.9730  5.6970 0.025410

DIC: 1193 
\end{verbatim}

\begin{figure}[h]
   \centering
   \includegraphics{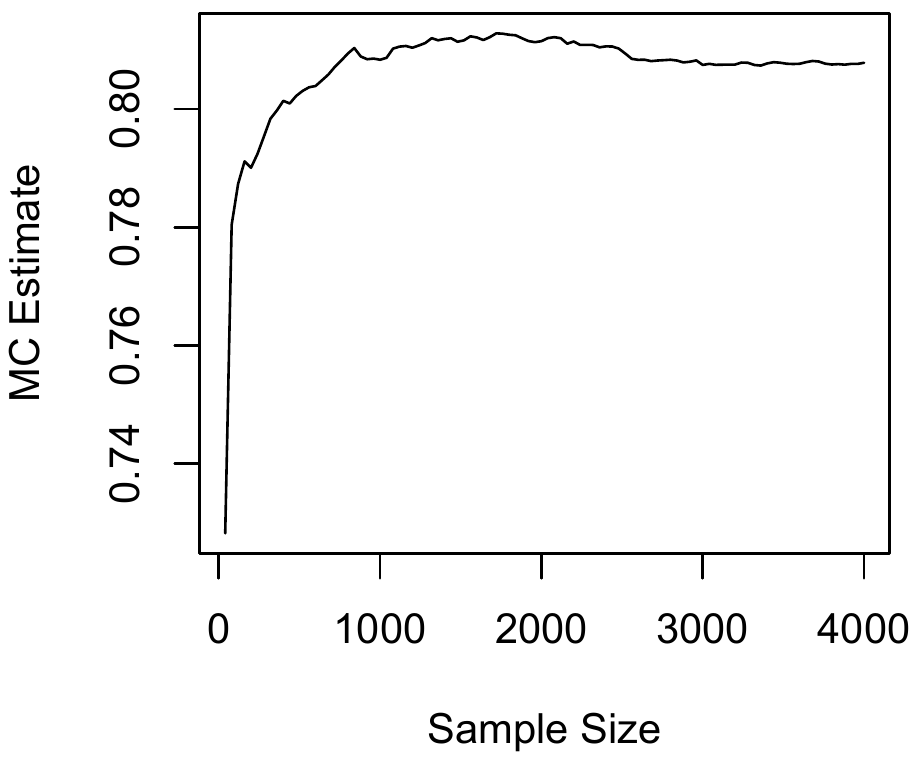}
   \caption{A plot of estimated posterior mean versus sample size for $\omega$, having assumed a Laplace marginal distribution.}
   \label{fig:trace}
\end{figure}

We see that sampling terminated when 4,000 samples had been drawn, since that sample size yielded $\widehat{\text{cv}}_j<0.01$ for $j\in\{1,2,3\}$. As a second check we examine the plot given in Figure~\ref{fig:trace}, which shows the estimated posterior mean for $\omega$ as a function of sample size. The estimate evidently stabilized after approximately 2,500 samples had been drawn.

The proposal standard deviations (1 for $\mu$, 0.1 for $\sigma$, and 0.2 for $\omega$) led to sensible acceptance rates of 40\%, 60\%, and 67\%.

\begin{verbatim}
R> fit1$accept

    inter        mu     sigma 
0.6694174 0.4013503 0.5951488 
\end{verbatim}

For a $t$ marginal distribution only 3,000 samples were required.

\begin{verbatim}
R> set.seed(4565)
R> fit2 = sklars.omega.bayes(data, verbose = FALSE,
+                            control = list(dist = "t", minit = 1000,
+                            maxit = 5000, tol = 0.01, sigma.1 = 0.2,
+                            sigma.2 = 2, sigma.omega = 0.2))
R> summary(fit2)

Call:

sklars.omega.bayes(data = data, verbose = FALSE, control = list(dist = "t", 
    minit = 1000, maxit = 5000, tol = 0.01, sigma.1 = 0.2, sigma.2 = 2, 
    sigma.omega = 0.2))

Number of posterior samples: 3000 

Control parameters:
                
dist        t   
minit       1000
maxit       5000
tol         0.01
sigma.1     0.2 
sigma.2     2   
sigma.omega 0.2 
                
Coefficients:

      Estimate   Lower  Upper     MCSE
inter    0.874  0.8283  0.919 0.002054
nu       6.720  5.0210  8.424 0.053200
mu      23.450 22.2600 24.690 0.028070

DIC: 1224 
\end{verbatim}

Note that the Laplace model yielded a much smaller value of DIC, and hence a very small relative likelihood for the $t$ model.

\begin{verbatim}
R> dic = c(fit1$DIC, fit2$DIC)
R> (pr = exp((min(dic) - max(dic)) / 2))

[1] 1.852924e-07
\end{verbatim}


\section{Summary and discussion}
\label{summary}

Sklar's $\omega$ offers a flexible, principled, complete framework for doing statistical inference regarding agreement. In this article we described three frequentist approaches to inference for Sklar's $\omega$ as well as a Bayesian methodology for interval or ratio outcomes. All of these approaches are supported by \proglang{R} package \pkg{sklarsomega}, version 2.0 of which is available from the Comprehensive \proglang{R} Archive Network. As illustrated in the preceding section, package \pkg{sklarsomega} also offers much useful related functionality.


\section*{Computational details}

The results in this paper were obtained using \proglang{R}~3.3.3 with the \pkg{extraDistr}~1.8.9 package, the \pkg{hash}~2.2.6 package, the \pkg{LaplacesDemon}~16.0.1 package, the \pkg{Matrix}~1.2.8 package, the \pkg{mcmcse}~1.3.2 package, the \pkg{numDeriv}~2014.2.1 package, the \pkg{spam}~1.3.0 package, and the \pkg{pbapply}~1.3.2 package. \proglang{R} itself and all packages used are available from the Comprehensive \proglang{R} Archive Network (CRAN) at \url{http://CRAN.R-project.org}.


\bibliography{refs}
\bibliographystyle{apa}

\end{document}